\documentclass[doublecol,figures]{epl2}
\usepackage{graphicx}
\usepackage{dcolumn}
\usepackage{bm}
\usepackage{amsmath}
\usepackage{amssymb}
\newcommand{\veck}{{\bf k}}
\newcommand{\veci}{{\bf i}}
\newcommand{\vecj}{{\bf j}}
\newcommand{\vecl}{{\bf l}}
\title{Variational study of Fermi-surface deformations in Hubbard models }
\author{J\"org B\"unemann\inst{1} \and Tobias Schickling\inst{2} \and Florian Gebhard\inst{2}}
\institute{
\inst{1} Institut f\"ur Physik, BTU Cottbus, D-03013 Cottbus, 
Germany\\
\inst{2}Fachbereich Physik, Philipps Universit\"at Marburg,
D-35032 Marburg, Germany}

\shortauthor{J.\ B\"unemann, T.\ Schickling, and F.\ Gebhard}


\abstract{We study the correlation-induced deformation of Fermi surfaces by
means of a new diagrammatic method which allows for the analytical evaluation 
of Gutzwiller wave functions in finite dimensions. In agreement with 
renormalization-group results we find Pomeranchuk instabilities
in two-dimensional Hubbard models for sufficiently large 
Coulomb interactions.}

\begin{document}
\maketitle
\section{Introduction}
The shape of the Fermi surface (FS) is one of the
most important properties that determine the low-energy 
physics of electron liquids.
The single-particle energy levels of non-interacting electrons
depend on the crystal momentum $\veck$ from the Brillouin zone
through the (single-band) dispersion relation
$\varepsilon(\veck)$. For $N$ electrons at zero temperature,
all single-particle states 
which lie below the Fermi energy $E_{\rm F}$ are occupied. 
The FS separates occupied and unoccupied single-particle levels, i.e.,
it consists of all $\veck$-points 
which obey the equation $E_{\rm F}-\varepsilon(\veck_{\rm F})=0$.
In the presence of finite Coulomb interactions, 
the calculation of the FS requires the real part of the 
proper self-energy $\Sigma(\veck,\omega)$ so that the FS is obtained from
\begin{equation}
E_{\rm F}-\varepsilon(\veck_{\rm F})-\mathfrak{Re}\Sigma(\veck_{\rm F},E_{\rm F})=0\;.
\end{equation}
Within perturbation theory for the Coulomb interaction~\cite{fetter2003},
the proper self-energy
is defined as the sum over all irreducible diagrams
for the single-particle Green function. 

The calculation of $\Sigma(\veck,\omega)$ 
is a notoriously difficult task in correlated-electron theory,
even for a single-band Hubbard model 
\begin{equation}
\hat{H}=\hat{H}_0 + U\sum_{\veci}  \hat{d}_{\veci}
\,, \, \hat{H}_0=\sum_{\veci,\vecj,\sigma}t_{\veci,\vecj}\hat{c}_{\veci,\sigma}^{\dagger}
\hat{c}_{\vecj,\sigma}^{\phantom{\dagger}} 
\, , \, \hat{d}_{\veci}\equiv\hat{n}_{\veci,\uparrow}\hat{n}_{\veci,\downarrow}
\label{eq:1}
\end{equation}
in two dimensions. Here, $\veci=(i_1,i_2)$ denotes one of 
the $L$~sites on a square lattice, and $\sigma=\uparrow,\downarrow$.  
The dispersion relation is given by 
\begin{equation}
\varepsilon(\veck)
=\frac{1}{L}\sum_{\veci,\vecj}\exp[{\rm i}(\veci-\vecj)\veck]t_{\veci,\vecj}\;.
\end{equation}
For small~$U$, the self-energy $\Sigma(\veck,\omega)$ may be calculated
from straightforward perturbation theory~\cite{schweitzer1991,zlatic1997,halboth1997},
or using renormalisation-group methods~\cite{halboth2000b,freire2008}.
When $U$ is of the order of the bare bandwidth~$W$, or larger,
only purely numerical methods such as 
Quantum-Monte Carlo~\cite{bulut1994,preuss1995} 
and Exact Diagonalisation~\cite{eder2011} are available
which still suffer from serious 
finite-size limitations~\cite{varney2009,eder2011}.
In view of these significant problems on the theoretical side, 
our understanding of experiments on two-dimensional Fermi surfaces, e.g., 
those of doped cuprates, is far from satisfactory.
Moreover, only reliable many-particle approaches permit a meaningful
comparison of measured and calculated Fermi surfaces that  
may reveal the correct form of $\hat{H}_0$ in~(\ref{eq:1}); 
for a recent overview, see, e.g., Ref.~\cite{eder2011}. 

In this work, we introduce a new analytical 
scheme to evaluate expectation values 
for Gutzwiller wave functions in~{\sl finite\/} spatial dimensions
in a controlled way.
By construction, the Gutzwiller approach provides 
the Fermi surface of quasi-particles in Landau's Fermi-liquid theory.
Therefore, the Gutzwiller wave function is an 
appropriate tool for the calculation of correlation-induced 
FS deformations at moderate interaction strengths, $U\approx W$. Unlike
 numerical schemes for the evaluation of Gutzwiller wave 
functions~\cite{gros1988,yokoyama1988,paramekanti2004,eichenberger2007,eichenberger2009} our approach does not suffer from finite-size limitations. It therefore provides us with a momentum-space resolution that is needed for the study
 of Fermi-surface deformations.  
     
\section{Evaluation of Gutzwiller wave functions}
The variational wave functions introduced by 
Gutzwiller~\cite{gutzwiller1963} 
for the single-band Hubbard model~(\ref{eq:1})
have the form 
\begin{equation}
|\Psi_{\rm G}\rangle=\hat{P}_{\rm G}|\Psi_0\rangle
=\prod\nolimits_{\veci}\hat{P}_{\veci;{\rm G}}|\Psi_0\rangle \; ,
\label{eq:1.2}                                                  
\end{equation}
where $|\Psi_0\rangle$ is a (normalised) single-particle 
product state and the
local `Gutzwiller correlator' is defined by
\begin{equation}
\hat{P}_{\veci;{\rm G}}=1-(1-g)\hat{d}_{\veci}\;.
\end{equation}
Here, $g\geq 0$ is a 
variational parameter which allows for the optimisation of the 
average number of doubly-occupied lattice sites. In this work, 
we will use the more convenient definition 
\begin{equation}
\hat{P}_{\veci}=\sum_{\Gamma}\lambda_{\Gamma}
|\Gamma \rangle_{\veci\,\veci}\! \langle \Gamma |
\end{equation}
for the local correlator.
It contains the variational parameters 
$\lambda_{\Gamma}$ for the four local states 
\begin{equation}
|\Gamma\rangle_{\veci}
\in \left\{|\emptyset\rangle_{\veci}, |\uparrow\rangle_{\veci}, 
|\downarrow\rangle_{\veci}, |\uparrow\downarrow\rangle_{\veci}\right\}
\end{equation}
for the empty, singly, or doubly occupied site~$\veci$.
 To keep notations simple, 
we assume  that the wave functions $|\Psi_{\rm G}\rangle$ and 
$|\Psi_0\rangle$ are translationally invariant and paramagnetic. The
corresponding derivation for more general wave functions is straightforward. 
 
In order to determine the expectation value of the Hamiltonian~(\ref{eq:1}) 
with respect to the wave function~(\ref{eq:1.2}) 
we need to evaluate ($\veci\neq\vecj$)
\begin{eqnarray}
\langle\Psi_{\rm G}|\Psi_{\rm G} \rangle 
&=&
\Bigl\langle \prod_{\vecl}\hat{P}^2_{\vecl}\Bigr\rangle_{0}\; ,
\label{eq:1.5}\\
\langle\Psi_{\rm G}|\hat{d}_{\veci}|\Psi_{\rm G} \rangle 
&=&
\Bigl\langle \hat{P}_{\veci} \hat{d}_{\veci}\hat{P}_{\veci}
\prod_{\vecl(\neq \veci)}\hat{P}^2_{\vecl}\Bigr\rangle_{0} \; ,
\label{eq:1.5b} 
\\
\langle\Psi_{\rm G}|
\hat{c}^{\dagger}_{\veci,\sigma}\hat{c}_{\vecj,\sigma}^{\phantom{\dagger}} 
|\Psi_{\rm G} \rangle
&=&
\Bigl\langle
\widetilde{c}^{\dagger}_{\veci,\sigma}\widetilde{c}_{\vecj,\sigma}^{\phantom{\dagger}}
\prod_{\vecl(\neq \veci,\vecj)}\hat{P}^2_{\vecl}\Bigr\rangle_{0}
\; , \label{eq:1.5c} 
\end{eqnarray}
where $\langle\ldots\rangle_{0}$ denotes expectation values 
with respect to $|\Psi_0\rangle$ and
$\widetilde{c}_{\veci,\sigma}^{(\dagger)}\equiv
\hat{P}_{\veci}\hat{c}_{\veci,\sigma}^{(\dagger)}\hat{P}_{\veci}$.
As we will show below, our diagrammatic expansion significantly simplifies 
if we set  
\begin{equation}
\hat{P}^2_{\vecl}=1+x\hat{d}_{\vecl}^{\rm HF} \; ,
\label{eq:1.6}
\end{equation}
where 
\begin{equation}
\hat{d}_{\vecl}^{\rm HF}\equiv 
\hat{n}^{\rm HF}_{\vecl,\uparrow}\hat{n}^{\rm HF}_{\vecl,\downarrow} \;\;,\;\;
\hat{n}^{\rm HF}_{\vecl,\sigma}\equiv\hat{n}_{\vecl,\sigma}-n_0\;,
\end{equation}
and $n_0\equiv\langle\hat{n}_{\vecl,\sigma}\rangle_0=N/(2L)$.
Equation~(\ref{eq:1.6}) determines three of the four 
parameters $\lambda_{\Gamma}$ as well as the coefficient~$x$.
In this way, we are left with only one variational parameter. 
For instance, we may express the parameters $\lambda_{\emptyset}$, 
$\lambda_{1}\equiv\lambda_{\sigma}$ for empty and singly-occupied sites
by $\lambda_d$, the parameter for doubly-occupied sites. 
Alternatively, due to the relations
\begin{equation}
x=[\lambda^2_d-1]/(1-n_0)^2 \;\Leftrightarrow 
\;\lambda^2_d=1+x(1-n_0)^2 \; ,
\label{eq:1.8}
\end{equation}
we may also consider~$x$ as our variational parameter. 
The expansion~(\ref{eq:1.6})
was first introduced in Ref.~\cite{gebhard1990} 
where it has been used as a convenient tool for the 
evaluation of expectation values in infinite dimensions~$D=\infty$ 
and of $1/D$~corrections. 
For another series expansion for the Gutzwiller 
 wave function around the limit $D=\infty$, see Ref.~\cite{metzner1989}. 
Note that both local correlators
$\hat{P}_{\veci;{\rm G}}$ and  $\hat{P}_{\veci}$, together with the 
condition (\ref{eq:1.6}),  
lead to the same variational space if the single-particle wave function 
$|\Psi_0\rangle$ is treated as a variational object.

With Eq.~(\ref{eq:1.6}), the norm and the expectation values 
Eqs.~(\ref{eq:1.5})--(\ref{eq:1.5c}) are given in form of a 
power series in~$x$, 
{\arraycolsep=2pt\begin{eqnarray}
\langle\Psi_{\rm G}|\Psi_{\rm G} \rangle 
&=&\sum_{k=0}^{\infty}\frac{x^k}{k!}
\sideset{}{'}\sum_{\vecl_1,\ldots \vecl_k}
\bigl\langle \hat{d}^{\rm HF}_{\vecl_1,\ldots,\vecl_k}
\bigr\rangle_{0}
\label{eq:1.9}\, ,\\
\langle\Psi_{\rm G}|\hat{d}_{\veci}^{\vphantom{\rm HF}}|\Psi_{\rm G} \rangle 
&=&\lambda_d^2\sum_{k=0}^{\infty}\frac{x^k}{k!}
\sideset{}{'}\sum_{\vecl_1,\ldots \vecl_k}
\bigl\langle \hat{d}_{\veci}^{\vphantom{\rm HF}}\hat{d}^{\rm HF}_{\vecl_1,\ldots,\vecl_k}
\bigr\rangle_{0}
\label{eq:1.9b}\, ,
\\
\langle\Psi_{\rm G}|\hat{c}^{\dagger}_{\veci,\sigma}\hat{c}_{\vecj,\sigma}^{\phantom{\dagger}} 
|\Psi_{\rm G} \rangle&=&
\sum_{k=0}^{\infty}\frac{x^k}{k!}
\sideset{}{'}\sum_{\vecl_1,\ldots \vecl_k}
\bigl\langle
\widetilde{c}_{\veci,\sigma}^{\dagger}
\widetilde{c}_{\vecj,\sigma}^{\phantom{\dagger}}
\hat{d}^{\rm HF}_{\vecl_1,\ldots,\vecl_k}\bigr\rangle_{0} \, ,
\label{eq:1.9c}
\end{eqnarray}}
where we introduced the notation
\begin{equation}
\hat{d}^{\rm HF}_{\vecl_1,\ldots,\vecl_k}\equiv\hat{d}^{\rm HF}_{\vecl_1}\cdots
\hat{d}^{\rm HF}_{\vecl_k}\quad, \quad
\hat{d}^{\rm HF}_{\emptyset}\equiv 1\;.
\end{equation}
The primes in Eqs.~(\ref{eq:1.9})--(\ref{eq:1.9c}) indicate the summation 
restrictions $\vecl_p\neq \vecl_{p'}$, $ \vecl_p \neq  \veci,\vecj$ 
for all $p,p'$. Note that the same expressions 
arise for the original Gutzwiller correlator $\hat{P}_{\veci;{\rm G}}$ 
when we replace $x$ and $\hat{d}^{\rm HF}_{\vecl}$ 
by $(g^2-1)$ and $\hat{d}_{\vecl}$, respectively~\cite{metzner1988,kollar2002}.
As we will demonstrate below, our expansion with respect to
$x$ converges significantly faster than the expansion in $(g^2-1)$.
Therefore, the first few orders in the $x$-expansion 
permit us to evaluate the Gutzwiller wave function accurately
for not too large interaction strengths.

The expectation values in Eqs.\ (\ref{eq:1.9})--(\ref{eq:1.9c}) 
can be evaluated by means of Wick's theorem~\cite{fetter2003}.
By construction, we eliminated all diagrams with local `Hartree bubbles' 
at internal vertices. 
To achieve the same 
for the external vertices we rewrite the corresponding operators 
in Eqs.~(\ref{eq:1.9b}), (\ref{eq:1.9c}) as
\begin{eqnarray}
\hat{d}_{\veci}&=&(1-xd_0)\hat{d}^{\rm HF}_{\veci}
+n_0(\hat{n}^{\rm HF}_{\veci,\uparrow}+\hat{n}^{\rm HF}_{\veci,\downarrow}) 
+d_0\hat{P}^2_{\veci}
\label{eq:1.10a}\; ,
\\
\widetilde{c}_{\veci,\sigma}^{(\dagger)}&=&
\hat{P}_{\veci}\hat{c}^{(\dagger)}_{\veci,\sigma}\hat{P}_{\veci}=
q \hat{c}^{(\dagger)}_{\veci,\sigma}
+\alpha\hat{c}^{(\dagger)}_{\veci,\sigma}\hat{n}^{\rm HF}_{\veci,\bar{\sigma}}
\label{eq:1.10b}\; ,
\end{eqnarray}
where we introduced
\begin{eqnarray}
d_0&\equiv& n_0^2\;,\\
q&\equiv&\lambda_1(\lambda_dn_0+\lambda_{\emptyset}(1-n_0))\;,\\
\alpha&\equiv&\lambda_1(\lambda_d-\lambda_\emptyset)\;,
\end{eqnarray}
 and 
$\bar{\uparrow}=\downarrow$, $\bar{\downarrow}=\uparrow$.
When inserted into~(\ref{eq:1.9b}), the last term in~(\ref{eq:1.10a}) 
combines to $\lambda_d^2 d_0\langle \Psi_{\rm G}|\Psi_{\rm G}\rangle$
so that it does not have to be evaluated diagrammatically.
In the resulting diagrammatic expansion of Eqs.~(\ref{eq:1.9})--(\ref{eq:1.9c}),
the $k$th-order terms correspond to diagrams 
with $k$ `internal' vertices on sites $\vecl_1,\ldots,\vecl_k$, one (two) 
`external' vertices on site $\veci$ ($\veci$ and $\vecj$) 
and lines 
\begin{equation}\label{pl}
P^{\sigma}_{\vecl,\vecl'}\equiv 
\langle\hat{c}^{\dagger}_{\vecl,\sigma}\hat{c}_{\vecl',\sigma}^{\phantom{\dagger}}\rangle_0
-\delta_{\vecl,\vecl'}n_0
\end{equation}
 connecting these vertices.

As the final analytical step of our derivation,  we apply 
the linked-cluster theorem~\cite{fetter2003}.
The norm~(\ref{eq:1.9}) cancels the disconnected diagrams 
in the two denominators~(\ref{eq:1.9b}) and~(\ref{eq:1.9c}). 
Note that a straightforward application of this theorem is hampered 
by the summation restrictions in these equation. However, applying
Wick's theorem to the expectation values in~(\ref{eq:1.9})--(\ref{eq:1.9c}) 
is equivalent to the evaluation of determinants such as 
$|P^{\sigma}_{\vecl,\vecl'}|$ with $\vecl,\vecl'\in (\vecl_1,\ldots,\vecl_k)$. 
Since these determinants vanish if any of the 
lattice sites $\vecl_1,\ldots,\vecl_k$ are the same, we can 
lift the summation restrictions in~(\ref{eq:1.9})--(\ref{eq:1.9c}) 
without creating additional terms. 

\begin{figure}[t] 
\onefigure[width=8cm]{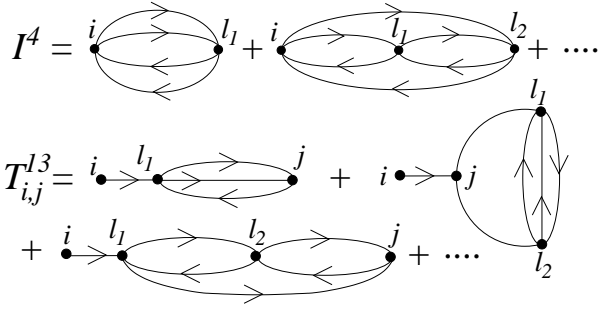} 
\caption{Diagrams with up to two internal vertices
for $I^{(4)}$ and $T_{\veci,\vecj}^{(1),(3)}$.
\label{Fig:fig1}}
\end{figure}

The remaining task is to evaluate the six diagrammatic sums
\begin{equation}
S=\sum_{k=0}^{\infty}\frac{x^k}{k!}S(k)
\end{equation}
with
\begin{equation}
S\in\{I^{(2)}, I^{(4)}, T^{(1),(1)}, T^{(1),(3)}, T^{(3),(1)}, T^{(3),(3)}\}
\end{equation}
and
{\arraycolsep=2pt\begin{eqnarray}\label{sd}
I^{(2)[(4)]}(k)&\equiv&\sum_{\vecl_1,\ldots, \vecl_k}
\bigl\langle 
\hat{n}^{\rm HF}_{\veci,\sigma}[\hat{d}^{\rm HF}_{\veci}]
\hat{d}^{\rm HF}_{\vecl_1,\ldots,\vecl_k}
\bigr\rangle^{\rm c}_{0}
\label{eq:urtz}\; ,\\
T_{\veci,\vecj}^{(1)[(3)],(1)[(3)]}(k)&\equiv&
\sum_{\vecl_1,\ldots,\vecl_k}
\bigl\langle
[\hat{n}^{\rm HF}_{\veci,\bar{\sigma}}]
\hat{c}^{\dagger}_{\veci,\sigma}
[\hat{n}^{\rm HF}_{\vecj,\bar{\sigma}}]
\hat{c}_{\vecj,\sigma}^{\phantom{\dagger}}\hat{d}^{\rm HF}_{\vecl_1,\ldots,\vecl_k}
\rangle^{\rm c}_{0}\;.
\nonumber
\end{eqnarray}}%
Here, $\langle \dots \rangle_0^{ \rm c}$ indicates that only connected diagrams 
are to be kept. As examples, we show the leading order diagrams of  
$I^{(4)}$ and  $T_{\veci,\vecj}^{(1),(3)}$ in Fig.~\ref{Fig:fig1}.

The variational ground-state energy functional is given by
\begin{equation}\label{355}
\langle \hat{H} \rangle_{\rm G}=E_0(|\Psi_0\rangle,x)=L(E^{\rm kin}+Ud)
\end{equation}
in the form 
\begin{eqnarray}\label{356}
\langle \hat{H} \rangle_{\rm G}
&=& 2\sum_{\veci,\vecj}t_{\veci,\vecj}
\big(q^2T_{\veci,\vecj}^{(1),(1)}
+2q\alpha T_{\veci,\vecj}^{(1),(3)}
+\alpha^2 T_{\veci,\vecj}^{(3),(3)}\big)
\nonumber\\
&&+LU \lambda_d^2\big((1-xd_0)I^{(4)}+2n_0I^{(2)}+d_0\big) \; ,
\label{eq:erz}
\end{eqnarray}
where $|\Psi_0\rangle $ enters the energy expression solely through the lines
$P^{\sigma}_{\veci,\vecj}$ and through $n_0$. 
In our case of a translationally invariant wave function, 
we have 
\begin{equation}\label{2se}
\langle \hat{n}_{\veci,\sigma} \rangle_{\rm G}=n_0\;.
\end{equation}
The l.h.s. of~(\ref{2se})  is given
 diagrammatically as
\begin{eqnarray}\label{2sep}
\langle \hat{n}_{\veci,\sigma} \rangle_{\rm G}&=&
\lambda^2_{d}\big(d_0+I^{(4)}(1-xd_0)+2n_0I^{(2)}\big)\\\nonumber
&&+\lambda^2_{1}
\big(m^0_{1}+I^{(2)}(1-2n_0)-I^{(4)}(1+x)m^0_{1}\big)\;,
\end{eqnarray}
where $m^0_{1}=n_0(1-n_0)$. From~(\ref{2se}), (\ref{2sep}), (\ref{eq:1.8}), 
and  
\begin{equation}
\lambda_1^2=1-n_0(1-n_0)x\;,
\end{equation}
 we then find
 the following relation between
 $I^{(4)}$ and $I^{(2)}$,
\begin{equation}
I^{(2)}=-I^{(4)} \frac{x(1-2n_0)}{1+xn_0(1-n_0)}\;.
\end{equation}
Therefore, only $I^{(4)}$ in~(\ref{eq:erz}) needs to be calculated. 

\section{The one-dimensional Hubbard model}
\begin{figure}[t]
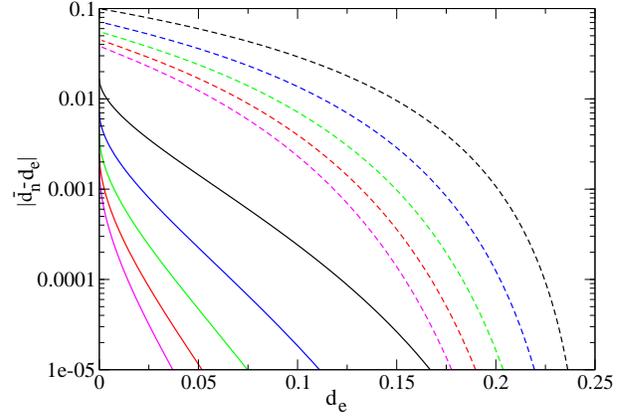
 
\onefigure[width=8cm]{pl2.eps} 
\caption{Difference between the
exact double occupancy $d_{\rm e}$  
and its $n$-th order Taylor expansions $\bar{d}_n$ 
(with $n=3,5,7,9,11$ in one dimension  (in descending order) 
with expansion parameters $x$ (solid line) and $(g^2-1)$ (dashed line).
\label{fig2}}
\end{figure}

We have tested the quality of our $x$-expansion against
the exact results for the Gutzwiller wave function
in one dimension~\cite{metzner1988,kollar2002}. From
the analytical results one can, e.g., determine 
the Taylor expansions of the average double occupancy 
with respect to~$x$ or $(g^2-1)$~\cite{gulacsi1993}. 
It turns out that the $x$-expansion converges much faster than the 
expansion in $(g^2-1)$. 
This can be seen, for example, in Fig. \ref{fig2} where 
the differences between the 
 exact double occupancy ($d_{\rm e}$)  and its $n$-th order Taylor expansions 
($\bar{d}_n$)  at half  band-filling ($n_0=1/2$) 
are shown for both parameters  as a 
function of  $d_{\rm e}$.
The figure reveals that the $x$-expansion to third order
is by an order of magnitude closer to the exact result 
than the $11$th-order expansion in $(g^2-1)$.

\begin{figure}[t]
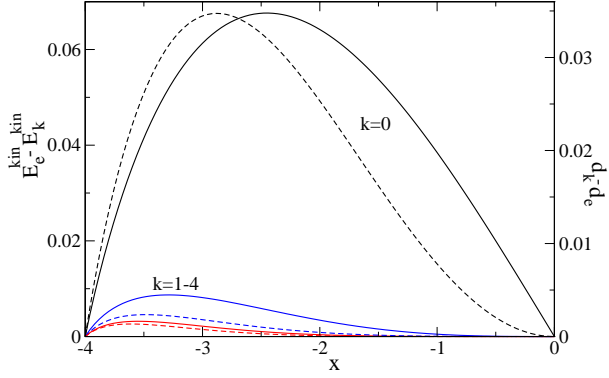
 
\onefigure[width=8.0cm]{fig3.eps} 
\caption{Differences between the
exact double occupancy $d_{\rm e}$ (right axis) and kinetic energy 
$E_{\rm e}^{\rm kin}$ (left axis)
within the Gutzwiller wave function 
in one dimension at half band-filling~\protect\cite{metzner1988,kollar2002}
and the corresponding $k$th order diagrammatic results 
for the double occupancy $d_k$ (solid lines, $n=0,1,3$,
in descending order) and the 
kinetic energy $E_k^{\rm kin}$ (dashed lines, $k=0,2,4$,
in descending order).
\label{Fig:fig3}} 
\end{figure}

In order to assess the absolute quality of our diagrammatic $x$-expansion,
we prefer to define the `order' of the expansion 
by the number of internal vertices which we retain in the diagrams. 
Consequently, the corresponding expressions for the
kinetic energy $E^{\rm kin}_k$ and the double occupancy $d_k$ to $k$th order
are {\sl not\/} identical to the $k$th order Taylor expansions
because the coefficients $q$, $\alpha$, $\lambda_d$ also depend on~$x$. 
In Fig.\ \ref{Fig:fig3} we show the deviations of the results to
leading orders ($k\le 4$) for the kinetic energy and the double occupancy 
from the analytic results in one dimension at half band-filling ($n_0=1/2$) 
as a function of $x$ ($x=-4$ corresponds to the atomic limit $d=0$). 
At half band-filling the lowest-order results ($k=0$) are equivalent to
the Gutzwiller approximation~\cite{metzner1988,gebhard1990}
because $\alpha(n_0=1/2)=0$. Moreover, the even (odd) orders $k>0$
vanish for the double occupancy (kinetic energy). 
Fig.~\ref{Fig:fig3} shows that the 4th-order results 
reproduce the exact results very well.

The one-dimensional Hubbard model is the worst case for 
our formalism because the latter is exact for the Gutzwiller wave function
in the opposite limit of infinite dimensions $D=\infty$
where all diagrams vanish. Therefore, 
we expect that the 4th-order results for the 
Gutzwiller wave function in two dimensions, which we discuss in the following,
 are also quite accurate. 

\section{Fermi-surface deformations in two-dimensional Hubbard models}
For the two-dimensional Hubbard model,
we treat $|\Psi_0\rangle$ and its FS as variational quantities.
The minimisation of~(\ref{eq:erz}) with respect to $|\Psi_0\rangle$ leads to 
the effective single-particle equation
\begin{eqnarray}\label{eq:iou}
\hat{H}_0^{\rm eff} |\Psi_0\rangle&=&
\sum_{\veci,\vecj}t^{\rm eff}_{\veci,\vecj}
\hat{c}_{\veci,\sigma}^{\dagger}\hat{c}_{\vecj,\sigma}^{\phantom{\dagger}}
|\Psi_0\rangle=
E^{\rm eff}|\Psi_0\rangle \;,\\\label{eq:iou1}
t^{\rm eff}_{\veci,\vecj}&=&
\frac{\partial E_0(|\Psi_0\rangle,x)}{\partial P^{\sigma}_{\veci,\vecj}} \; .
\end{eqnarray}
The effective dispersion relation
\begin{equation}
\varepsilon^{\rm eff}(\veck)
=\frac{1}{L}\sum_{\veci,\vecj}\exp^{{\rm i}(\veci-\vecj)\veck}
t^{\rm eff}_{\veci,\vecj}
\label{eq:effedispersion}
\end{equation}
defines  the quasi-particle FS via
\begin{equation} 
\varepsilon^{\rm eff}(\veck_{\rm F})=E_{\rm F}\; 
\label{eq:FS}
\end{equation}
because the correlated momentum distribution
 \begin{equation} 
n_{\veck,\sigma}\equiv \langle \hat{c}^{\dagger}_{\veck,\sigma}\hat{c}^{}_{\veck,\sigma} \rangle_{\rm G}
\end{equation}
 has step discontinuities exactly at the momenta given by Eq.~(\ref{eq:FS}).

The remaining problem is to solve self-consistently the 
closed set of equations~(\ref{355}), (\ref{356}), 
(\ref{eq:iou})-(\ref{eq:FS}), 
together with 
\begin{equation}
\frac{\partial}{\partial x} E_0(|\Psi_0\rangle,x) =0\;.
\end{equation}
Note that $E^{\rm eff}$  is just an auxiliary quantity, the 
ground-state energy of the effective single-particle 
Hamiltonian $\hat{H}_0^{\rm eff}$ . 
It must not be confused with the variational ground-state energy~(\ref{356}).
Numerically, we determine
 $|\Psi_0\rangle $ by solving~(\ref{eq:iou})  in momentum space while the  
diagrammatic sums in~(\ref{356}) and the derivatives in~(\ref{eq:iou1}) are 
evaluated in position space (up to 4th~order in this work).  For example, 
in a paramagnetic state, the 
 first two contributions of $I^4$ in Fig.~\ref{Fig:fig1} are given by
\begin{equation}\label{qwe}
I^4=x\sum_{l_1}P^4_{i,l_1}+4\frac{x^2}{2}
\sum_{l_1,l_2}P^2_{i,l_1}P^2_{i,l_2}P^2_{l_1,l_2}+
\cdots
\end{equation} 
where the lines~(\ref{pl}) were assumed to be 
spin independent, $P_{i,j}\equiv P^{\sigma}_{i,j}$. Note that the combinatorial 
factor $4$ in~(\ref{qwe}) 
results from the different possibilities to label the 
 lines with spin indices $\sigma$. We have determined these factors 
by means of a computer algorithm because their calculation 
becomes quite involved for higher-order diagrams. In this context, 
it was particularly helpful to use the exact results in one dimension
to eliminate programming errors.
  
\begin{figure}[t] 
\includegraphics[width=8.3cm]{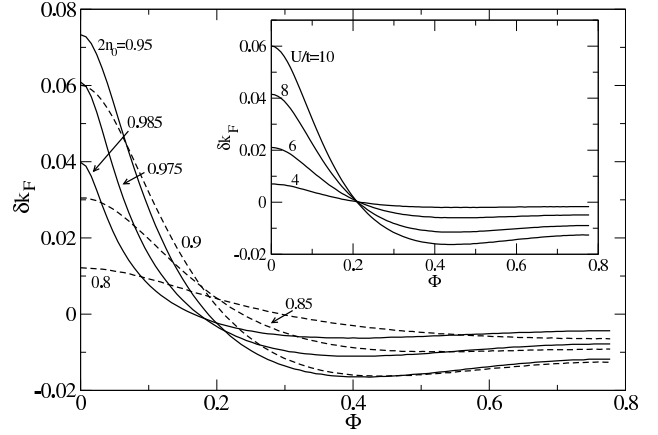} 
\caption{Polar plot of the FS deformations $\delta k_{F}$ 
 for $U/t=10$ and various band fillings $2n_0$; Inset: 
 FS deformations $\delta k_{F}$ for $2n_0=0.9$ and various values
 of $U/t$.}
\label{fig4b} 
\end{figure}

In principle, the effective hopping parameters 
$t^{\rm eff}_{\veci,\vecj}\equiv t^{\rm eff}_{X,Y}$
have to be calculated 
for arbitrary distances $X\equiv i_1-j_1$, $Y\equiv i_2-j_2$
of the lattice vectors $\veci=(i_1,i_2)$ and $\vecj=(j_1,j_2)$.
However, since the calculation of the derivatives in~(\ref{eq:iou1})
is numerically quite expensive, we take into account only 
the dominant parameters $t^{\rm eff}_{X,Y}$. In our  
calculations we include the seven 
hopping parameters with $X^2+Y^2\le 10$. To be consistent,
we also set $P^{\sigma}_{X,Y}=0$ for all $X^2+Y^2 > 10$ in  
the energy functional $E_0(|\Psi_0\rangle,x)$. 
This simplifies the numerical evaluation of the diagrams and it 
ensures that the self-consistency equations~(\ref{eq:iou1})
lead to a (local) minimum of $E_0(|\Psi_0\rangle,x)$. Note that the 
numerical calculation of all diagrammatic sums in~(\ref{356}) takes
 less than a minute on a present-day desktop computer. Once evaluated,
 the variational ground-state energy can be calculated immediately 
for {\sl any}  value of $U$. This illustrates that the study  
 of more complicated wave functions or multi-band 
 models in two (or even three) dimensions is a feasible problem within
 our approach.

\begin{figure}[t]
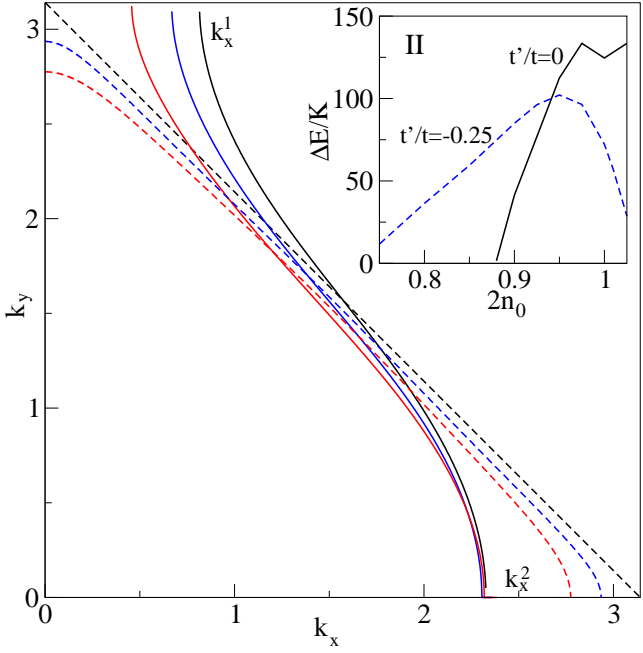
 
\onefigure[width=8.5cm]{pom2.eps}
\caption{Pomeranchuk FS for the
Hubbard model with nearest-neighbour hopping $t_{1,0}=-t$ 
for $U/t=10$ (solid) in comparison with the FS for $U/t=0$ (dashed)
for $2n_0=1.0,0.95,0.9$ (from right to left). 
Inset: energy gain $\Delta E$ in Kelvin ($t=1.0\, {\rm eV}$)
in the symmetry-broken phase as a function of the filling $2n_0$
for $U/t=10$.
\label{Fig:pom2a}} 
\end{figure} 

We first consider a Hubbard model with only 
 nearest-neighbour hopping $t_{1,0}=-t$  on a square lattice.
Since in this case the FS deformations 
$\delta k_{F}\equiv k_{F}(U)- k_{F}(0)$ are  rather small we plot them 
in a polar diagram as a function of the 
angle $\Phi=\arctan{(k_y/k_x)}$. Due to the symmetry of the lattice, 
we only need to 
consider angles $\Phi\in (0,\pi/4)$.  In Fig.~\ref{fig4b} we show  
$\delta k_{F}$ as a function of $\Phi$ for various hole-dopings 
$2n_0<1$ near half filling. 
Due to particle-hole symmetry, it is sufficient to study
hole dopings and the FS is unchanged at half band-filling. 
At finite doping, the overall feature of the FS deformation has a 
maximum for $2n_0\approx 0.95$ and it becomes negligible for densities 
$2n_0<0.75$. Note that the $\Phi$-dependence  of $\delta k_{F}$  is a 
non-trivial function of the doping.  For example, the two curves with 
$2n_0=0.9$ and  $2n_0=0.975$ have almost the same deformation at 
$\Phi=0$ (i.e., in [0,1] direction) but differ significantly for finite 
$\Phi$. In contrast, 
for fixed $n_0$, only the slope of the curves becomes 
smaller when we decrease $U$; see the inset of Fig.~\ref{fig4b}.

For sufficiently large values of $U/t$ and close to half band-filling, 
these Fermi surfaces are unstable against 
the Pomeranchuk 
effect~\cite{pomeranchuk1958,halboth2000,yamase2000,yamase2000b}, i.e.,
we find minima in the variational space which break 
the discrete $C_{4}$ symmetry of the lattice,
in agreement with related 
numerical work on $t$-$J$ models~\cite{edegger2006b,yamase2007,jedrak2010}.
In Fig.~\ref{Fig:pom2a} we show the FS for $2n_0=0.9,0.95,1.0$
for $U/t=10$  (Pomeranchuk phases) and $U/t=0$.
Around half band-filling, $|2n_0-1|\lesssim 0.12$, the states 
with broken symmetry are stable. As seen from the inset
of Fig.~\ref{Fig:pom2a}, the maximal energy gain due to the symmetry 
breaking is of a fraction of room temperature,
$\Delta E/k_{\rm B}\approx 100\, {\rm K}$ for $t=1.0\, {\rm eV}$.

The asymmetry of the Pomeranchuk FS at finite doping is 
quite remarkable. The two intersection points, denoted as $k_x^1$ and 
$k_x^2$ in Fig.~\ref{Fig:pom2a}, obey $k_x^1=\pi-k_x^2$ 
only at half band-filling.
The Pomeranchuk minima are not continuously connected 
to those without broken symmetries. 
Therefore, in our approach, 
the transitions between such states as a function of $U/t$
are discontinuous, in general.

\begin{figure}[t]
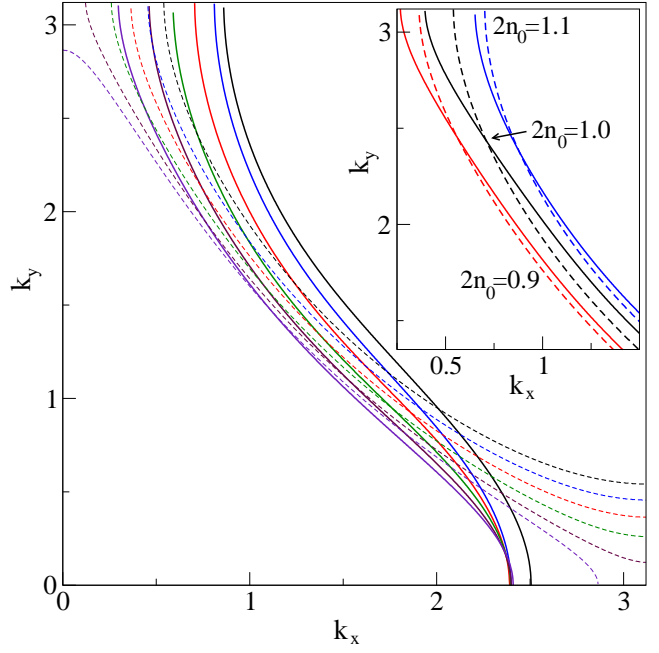
 
\onefigure[width=8.5cm]{pl3b.eps}
\caption{Pomeranchuk FS for the
Hubbard model with nearest-neighbour hopping $t_{1,0}=-t$
and second-neighbour hopping $t_{1,1}=-t'=0.25t$
for $U/t=10$ (solid) in comparison with the FS for $U/t=0$ (dashed)
for $2n_0=0.75,\ldots,1.0(0.05)$ (from left to right). Inset:  
FS for $U/t=10$ (solid) and $U=0$ (dashed) 
for $2n_0=0.9,1.0,1.1$ (from left to right) 
in the lattice-symmetric phase.\label{Fig:pom2b}} 
\end{figure} 

Second, we consider the Hubbard model 
with an additional second-neighbour hopping $t_{1,1}=-t'=0.25t$.
Even in the absence of symmetry breaking,
the FS deformations are much larger than in the 
Hubbard model with nearest-neighbour hopping only, 
as seen from the inset of Fig.~\ref{Fig:pom2b} 
where we show the lattice-symmetric FS for $U/t=10$ and $U/t=0$
near half band-filling.   
The Pomeranchuk instabilities occur mainly for hole doping,
$0.70 \lesssim 2n_0\lesssim 1.03$ for $U/t=10$, see
the inset of Fig.~\ref{Fig:pom2a}.
In Fig.~\ref{Fig:pom2b}, we show 
the Pomeranchuk FS for $U/t=10$ and the corresponding FS for $U/t=0$ 
for densities $0.75 \leq 2n_0\leq 1$. 

Note that the results presented in Figs.~\ref{fig4b}-\ref{Fig:pom2b} 
are certainly altered by the inclusion of  
other symmetry-broken phases with, 
e.g., magnetic or superconducting order. Such phases will be investigated in 
future studies. {One should also keep in mind that our approach only allows 
 us to study states which are adiabatically connected to some non-interacting 
 reference system (`Fermi liquids'). This excludes, in particular,
 the investigation  of Mott-Hubbard insulators. At zero temperature, however,
 such insulators are usually found only in theoretical studies which 
deliberately neglect
 ordered states, such as antiferromagnetic spin waves.  

\section{Conclusions}
In summary, we have introduced a novel scheme for the evaluation of
Gutzwiller wave functions in finite dimensions.
As a first application, we described the
correlation-induced Fermi surface 
deformations in two-dimensional Hubbard models within the Gutzwiller approach.
To the best of our knowledge, there exists no alternative method to 
calculate quasi-particle Fermi surfaces
for interaction parameters $U \approx W$ in the Hubbard model.
Our approach can be extended in various directions:
to study magnetic and superconducting order
in single and multi-band Hubbard models~\cite{buenemann1998},
and to calculate dynamical response 
functions~\cite{seibold2001,buenemann2011b,buenemann2011f}.

\end{document}